\documentclass[sigconf]{aamas}

\usepackage{balance} % for balancing columns on the final page
\usepackage{amsmath}
\usepackage{algorithm}
\usepackage{mathtools}
\usepackage{mathrsfs}
\usepackage{bm}
\usepackage{framed}
\usepackage{mdframed}
\usepackage{lipsum}
\usepackage{algpseudocode}
\usepackage{mathtools}
\usepackage{mathrsfs}
\usepackage{graphicx} 
\usepackage{comment}
\usepackage{balance} % for balancing columns on the final page
\usepackage{appendix}
\usepackage{subfigure}
\usepackage{stfloats}
\usepackage{comment} 
\usepackage{graphicx} 
\makeatletter
\gdef\@copyrightpermission{
  \begin{minipage}{0.2\columnwidth}
   \href{https://creativecommons.org/licenses/by/4.0/}{\includegraphics[width=0.90\textwidth]{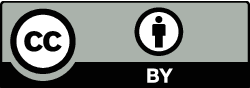}}
  \end{minipage}\hfill
  \begin{minipage}{0.8\columnwidth}
   \href{https://creativecommons.org/licenses/by/4.0/}{This work is licensed under a Creative Commons Attribution International 4.0 License.}
  \end{minipage}
  \vspace{5pt}
}
\makeatother

\setcopyright{ifaamas}
\acmConference[AAMAS '25]{Proc.\@ of the 24th International Conference
on Autonomous Agents and Multiagent Systems (AAMAS 2025)}{May 19 -- 23, 2025}
{Detroit, Michigan, USA}{Y.~Vorobeychik, S.~Das, A.~Nowé  (eds.)}
\copyrightyear{2025}
\acmYear{2025}
\acmDOI{}
\acmPrice{}
\acmISBN{}

\acmSubmissionID{<<fp0207>>}

\title[AAMAS-2025 Formatting Instructions]{Nucleolus Credit Assignment for Effective Coalitions in Multi-agent Reinforcement Learning}

\author{Yugu Li}
\affiliation{
  \institution{University of South Australia}
  \city{Adelaide}
  \country{Australia}}
\email{liyyy301@mymail.unisa.edu.au}

\author{Zehong Cao}
\authornote{*Corresponding author}
\affiliation{
  \institution{University of South Australia}
  \city{Adelaide}
  \country{Australia}}
\email{jimmy.cao@unisa.edu.au}

\author{Jianglin Qiao}
\affiliation{
  \institution{University of South Australia}
  \city{Adelaide}
  \country{Australia}}
\email{jianglin.qiao@unisa.edu.au}

\author{Siyi Hu}
\affiliation{
  \institution{University of South Australia}
  \city{Adelaide}
  \country{Australia}}
\email{siyi.hu@unisa.edu.au}

\begin{abstract}
In cooperative multi-agent reinforcement learning (MARL), agents typically form a single grand coalition based on credit assignment to tackle a composite task, often resulting in suboptimal performance. This paper proposed a nucleolus-based credit assignment grounded in cooperative game theory, enabling the autonomous partitioning of agents into multiple small coalitions that can effectively identify and complete subtasks within a larger composite task. Specifically, our designed nucleolus Q-learning could assign fair credits to each agent, and the nucleolus Q-operator provides theoretical guarantees with interpretability for both learning convergence and the stability of the formed small coalitions. Through experiments on Predator-Prey and StarCraft scenarios across varying difficulty levels, our approach demonstrated the emergence of multiple effective coalitions during MARL training, leading to faster learning and superior performance in terms of win rate and cumulative rewards especially in hard and super-hard environments, compared to four baseline methods. Our nucleolus-based credit assignment showed the promise for complex composite tasks requiring effective sub-teams of agents.
\end{abstract}


\newcommand{\BibTeX}{\rm B\kern-.05em{\sc i\kern-.025em b}\kern-.08em\TeX}

\begin{document}

%%% The following commands remove the headers in your paper. For final 
%%% papers, these will be inserted during the pagination process.

\pagestyle{fancy}
\fancyhead{}

%%% The next command prints the information defined in the preamble.

\maketitle 

%%%%%%%%%%%%%%%%%%%%%%%%%%%%%%%%%%%%%%%%%%%%%%%%%%%%%%%%%%%%%%%%%%%%%%%%

\section{Introduction}

In multi-agent environments, determining the contribution of each agent from reward signals is critical for effective cooperation to complete a composite task ~\cite{farinelli2020decentralized, DBLP:conf/atal/KimS0T22, aziz2023possible}. In cooperative multi-agent reinforcement learning (MARL), this process, known as Credit Assignment ~\cite{gronauer2022multi}, is central to improving both the performance and interpretability of MARL systems. Credit assignment approaches for MARL can be broadly classified into two categories: implicit and explicit methods.

Implicit credit assignment methods, such as VDN ~\cite{sunehag2017value}, QMIX ~\cite{rashid2020monotonic}, and WQMIX ~\cite{rashid2020weighted}, rely on value decomposition to indirectly infer the contribution of each agent from the learning process. While these methods scale efficiently, their reliance on predefined value decomposition structures often leads to weak performance, when the task environment does not align well with these assumptions. Moreover, they have limited interpretability, making it difficult to understand the specific contributions of individual agents during task execution ~\cite{chen2023learning}.

Explicit credit assignment methods, on the other hand, directly evaluate each agent’s contribution based on its actions, providing better interpretability and robust performance. Benchmark algorithms in this category include COMA ~\cite{foerster2018counterfactual}, SQDDPG ~\cite{wang2020shapley}, and SHAQ ~\cite{wang2022shaq}. COMA, an early approach, employs a centralized critic with a counterfactual baseline to assess each agent's action value. However, its ability to accurately attribute contributions from individual agents becomes restricted in complex environments. More recent methods, such as SQDDPG and SHAQ, leverage Shapley value, a concept from cooperative game theory, to fairly distribute rewards based on each agent’s contribution. These methods offer strong theoretical convergence guarantees and improved interpretability by evaluating the individual contributions of agents ~\cite{heuillet2021explainability}.

However, both implicit and explicit credit assignment methods predominantly assume that all agents form a single grand coalition to tackle a composite task. While promoting fairness and stability, they occasionally lead to inefficiencies, especially in real-world applications that involve multiple, smaller sub-teams of agents working on different subtasks. For instance, tasks such as resource management and emergency response (e.g., fire rescue) require agents to form smaller, dynamic coalitions that can cooperate on subtasks while still contributing to the overall mission ~\cite{bansal2017emergent,campbell2013multiagent,iqbal2022alma}. But existing credit assignment fails to incentivize the formation of these smaller coalitions, resulting in larger search spaces and reduced learning efficiency ~\cite{proper2009solving}.
\begin{figure*}[t]
    \centering
\includegraphics[width=0.95\linewidth]{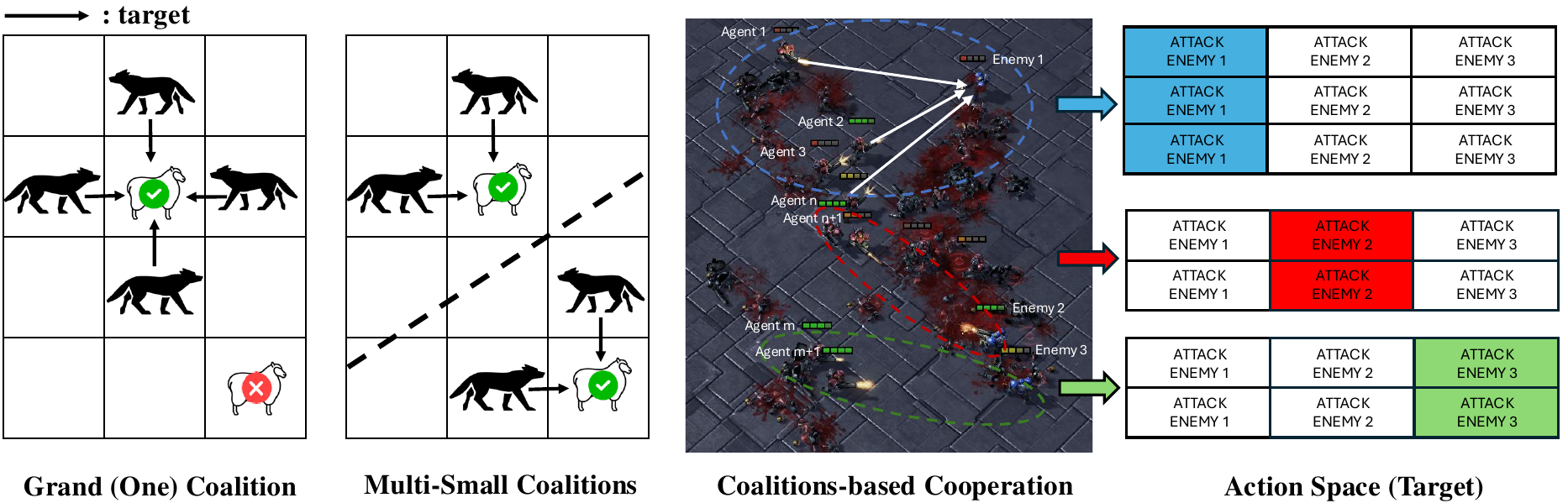}
    \caption{The transition from a single grand coalition to multiple smaller, task-specific coalitions in MARL. In scenarios like SMAC in super-hard maps: where a large number of agents are involved, forming multiple small coalitions is crucial for task completion efficiently. Agents who attack the same enemy unit naturally form these coalitions, enabling them to work together efficiently to achieve the mission. }
    \label{w-s}
\end{figure*}
To address this challenge, it is essential to enable the emergence of smaller, effective coalitions for cooperative MARL. We expect these small coalitions can only cooperate on subtasks, improving overall composite task performance and learning efficiency, as shown in \textbf{Fig. \ref{w-s}}. Agents can transit from a single grand coalition to multiple smaller, task-specific coalitions in MARL, and ensure that these smaller coalitions remain fair, interpretable, and stable over time.

In this paper, we propose a novel credit assignment approach based on the nucleolus concept from cooperative game theory ~\cite{schmeidler1969nucleolus}. The nucleolus is designed to minimize the maximum dissatisfaction among agents by providing fair reward allocations across multiple small coalitions rather than a single grand coalition. \textit{Fairness} is achieved by calculating each agent’s contribution relative to what each coalition could achieve independently, ensuring that no agent feels undervalued. In doing so, the nucleolus encourages agents to remain committed to their coalitions. Without fair reward allocation, agents may become dissatisfied and leave their coalitions. Nucleolus also improves \textit{interpretability} by explicitly linking reward allocation to the agents’ contributions, making it easier to trace why a particular coalition receives a given reward. \textit{Stability} is guaranteed as the nucleolus works to minimize excess—the gap between a coalition's potential independent gain and its current reward—thereby ensuring that no coalition has the incentive to deviate or reorganize, maintaining consistent cooperation throughout the task. Thus, our introduced nucleolus-based credit assignment for MARL will form stable and fair small coalitions. These coalitions partition a composite task into manageable subtasks, significantly improving both learning efficiency and MARL system performance. Additionally, we provide theoretical guarantees for the convergence of the proposed nucleolus-Q operator, ensuring that the action values of agents within each coalition converge to a local optimum. Through experiments on standard benchmarks such as Predator-Prey and StarCraft Multi-Agent Challenge, we demonstrate that our approach outperforms the existing four methods in terms of learning efficiency and overall performance.

%%%%%%%%%%%%%%%%%%%%%%%%%%%%%%%%%%%%%%%%%%%%%%%%%%%%%%%%%%%%%%%%%%%%%%%%

\section{Related Work}

\subsection{Credit Assignment in CTDE}

\paragraph{\textbf{Implicit Credit Assignment}}

Credit assignment is a fundamental challenge in Centralized Training with Decentralized Execution (CTDE) ~\cite{oliehoek2008optimal}. VDN ~\cite{sunehag2017value} introduces an additivity assumption, where the global Q-value is assumed to be the sum of the local Q-values of individual agents. This decomposition allows for simpler representation but significantly limits the expressiveness of the global Q-value. To overcome this issue, QMIX ~\cite{rashid2020monotonic} extends VDN by developing a mixing network that allows non-linear combinations of local Q-values to form the global Q-value. This non-linearity enhances the representative capacity of the global Q-value but remains limited by its implicit credit assignment mechanism, where agent contributions are not explicitly modeled. Consequently, QMIX struggles with environments featuring non-monotonic rewards.
WQMIX ~\cite{rashid2020weighted} further looks at this issue by introducing a learnable weighting mechanism to dynamically adjust the contributions of each agent’s local Q-value in the global Q-value computation. While WQMIX improves adaptability, the selection of weighting coefficients often relies on trial and error, with minimal interpretability. Moreover, improper weight adjustments can lead to instability in the learning process, further complicating its application in complex multi-agent environments.

\paragraph{\textbf{Explicit Credit Assignment}}

For explicit credit assignment methods, they can directly compute the contribution of each individual agent, ensuring that each agent's policy is optimized locally. One such method is COMA ~\cite{foerster2018counterfactual}, which discovers a counterfactual baseline to evaluate the difference in global reward when an agent takes alternative actions. COMA calculates each agent’s contribution to the current action by observing how global Q-value changes, but it suffers from inefficiency issues. Because the counterfactual baseline is derived from the current joint action rather than the optimal one, this credit assignment may not update to an optimal joint policy. Additionally, the counterfactual calculation in COMA increases computational complexity and reduces sample efficiency. More recently, methods such as Rollout-based Shapley Values ~\cite{10624777}, SQDDPG ~\cite{wang2020shapley}, and SHAQ ~\cite{wang2022shaq} have leveraged the Shapley value ~\cite{shapley1953value} from cooperative game theory to improve the interpretability of credit assignment in MARL. Shapley value-based methods offer a principled way to fairly distribute rewards based on each agent’s contribution. While SQDDPG builds on the DDPG network architecture and adopts a centralized approach similar to COMA, it is difficult for multi-agent Q-learning. SHAQ employs a decentralized approach, allowing for more scalable credit assignment, but these Shapley value-based methods assume a single grand coalition. In scenarios that require multi-agent cooperation based on multiple, smaller coalitions, the aforementioned explicit credit assignment methods do not achieve efficiency.

\subsection{Nucleolus in Cooperative Games}

In game theory, the Nucleolus concept ~\cite{schmeidler1969nucleolus} is a solution to identify the most equitable allocation of resources or payoffs in a cooperative game by minimizing the maximum dissatisfaction (or excess) among coalitions. For example, Gautam et al. ~\cite{10078455} highlight the role of the nucleolus in power and energy systems, emphasizing its effectiveness in resolving resource allocation, demand response, and system reliability issues. Similarly, Oner \& Kuyzu ~\cite{ONER2022108583} proposes nucleolus-based cost allocation methods and develops column and row generation techniques to solve the constrained lane covering game for optimizing truckload transportation procurement networks, which uses a nucleolus approach to minimize costs and fairly allocate them among shippers. These approaches are generally tailored for static reward settings with well-understood environments, rather than for maximizing rewards in MARL contexts. In dynamic environments, where rewards are delayed, traditional nucleolus-based methods fall short, necessitating the development of more flexible exploration policies for agents to adapt to uncertain conditions.

\section{Method}
We begin by developing an entity-based, partially observable coalition Markov decision process that supports coalition formation in Subsection 3.1. Next, our proposed nucleolus-based credit assignment in Subsection 3.2: (i) define the nucleolus-Q value by incorporating the concept of the nucleolus into Q-learning to ensure both the optimal coalition structure and optimal actions, supported by a theorem proof, and (ii) introduce a new nucleolus-based Bellman operator that converges to the optimal nucleolus Q-value, also with a corresponding theorem proof.

\subsection{Entity-based Coalition POMDP}
We extend Decentralized Partially Observable Markov Decision Process (Dec-POMDP) framework~\cite{iqbal2021randomized,liu2021coach,oliehoek2012decentralized} by defining an Entity-based Coalition POMDP, named as EC-POMDP. Here, entities include both controllable agents and other environment landmarks, and it is essential to distinguish the terms between a composite task and its subtasks: a subtask represents the smallest unit of reward feedback in the environment and cannot be further divided, whereas the collection of all subtasks is referred to as a composite task. Formally, the EC-POMDP framework is represented as $\langle S, N, A, P, O,\gamma, E, G, (E_g, A_g, r_g)_{g\in G}, R, CS \rangle$, where $S$ denotes a set of states of the environment; $N =\{1,\dots,n\}$ represents a set of agents; $A = \times_{i \in N} A_i$ is a set of joint actions of agents, each joint action $a=(a_1,\dots, a_n)\in A$, where $a_i$ represents the action of agent $i\in N$; $P$ is the state transition function defined as $P(s'|s, a)$; $o_i \in O$ is the agent's observation; $\gamma$ represents the learning rate;  $E$ is set of entities, where each entity $e \in E$ has state $s^{e} \in S$ and let $S^e\subseteq S$ define all possible states of entity $e$; $G=\{g_1,\dots,g_m\}$ is the composited task of the environment with $m$ subtasks; each subtask $g$ is defined as tuple $(E_g, A_{g},r_g)_{g\in G}$, where $E_g\subseteq E$ is a set of subtask-specific entities with their states $S^{E_g}=\bigcup_{e\in E_g}S^e$, $A_g\subseteq A$ representing actions used to complete subtasks $g$ and $r_g : S^{E_g} \times S \times A_g \to \vmathbb{R}$ is the reward function to complete sub-task $g$; $R$ is the total reward for the composite task $G$; $CS=\{C_g\subseteq N | \forall g\in G\}$ is the coalition structure, which satisfied with $\bigcup_{g\in G}C_g =N$ and $C_g\cap C_{g'}=\emptyset$ for all $g,g'\in G$, and each $C_g=\{i |\forall i\in N,s.t. a_i\in A_g\}$ is a coalition of agents that actions of agents is needed for each subtask $g$.

EC-POMDP is a Markov decision process that evolves continuously over discrete time steps. We assume that, in the initial state, all agents are in a single grand coalition status. As subtasks are identified during the process, this grand coalition breaks into multiple smaller coalitions, each dedicated to completing specific subtasks. Upon completing these subtasks, these smaller coalitions will reconsolidate into a grand coalition to complete the composite task finally.

\subsection{Nucleolus-based Credit Assignment}
%We begin by modeling the concept of the nucleolus and the Markov core in cooperative games into MARL, presented as the Nucleolus Q-value and summarized in Theorem 1. Following this, we introduce a novel nucleolus-based Bellman operator, designed to compute the optimal Nucleolus Q-value, termed Nucleolus Q-learning, and summarized in Theorem 2.
\subsubsection*{\textbf{Nucleolus in Cooperative Game}}
 
The nucleolus is a concept of fairness used to solve payoff distribution in cooperative games. It focuses on allocating payoffs among coalitions to minimize the maximum dissatisfaction (excess), thereby achieving a stable and fair distribution ~\cite{schmeidler1969nucleolus}. It provides stability and predictability, as it is a unique and definitive solution in every game ~\cite{rapoport2012game}. The nucleolus is typically located within or near the core of the game, ensuring the stability of the distribution scheme, and it balances the interests of different coalitions to reduce the likelihood of unfair allocation. This approach is widely applied in fields such as economics, politics, and supply chain management to benefit distribution in cooperation settings.

Let $(N,u,v)$ be a cooperative game ~\cite{branzei2008models}, where $N=\{1,\dots,n\}$ is the set of agents, $u(C)$ is the unity function to measure the profits earned by a coalition $C\subseteq N$, $v$ represents the total assets that can be distributed. Let $x=(x_1,\dots,x_n)$ be a payoff distribution over $N$. For a coalition $C$, the excess $e(C,x)$ of the coalition at $x$ is defined as
\begin{equation}e(C, x)=u(C)-\sum_{i \in C} x_i\end{equation}
where $x_i$ represented the payoff of agent$_i$ in coalition $C$. Nucleolus is a distribution method to minimize the maximum degree of dissatisfaction of coalition in cooperative games. Formally, the optimized process of the nucleolus is as follows:
\begin{equation}\min_{x} \max_{C\subseteq N} e(C, x)\end{equation}
Since there are $n$ agents, there are a total of $2^{n}$ coalitions (including empty sets), so for a set of imputation $x$, there are always $2^{n}$ excesses. And then sort excesses by non-increment called excesses sequence $\theta(x)$, and it is defined as follows:
\begin{equation}
\theta(x)=\left[e\left(C_1, x\right), e\left(C_2, x\right), \ldots, e\left(C_{2^n}, x\right)\right]
\end{equation}
The final nucleolus $x^*$ is a unique payoff allocation that satisfies:
\begin{equation}x^*=\{\theta(x)\preceq\theta(y)\mid \forall y \in I(v)\}\end{equation}
where $\preceq$ is lexicographical order and $I(v)$ is the set of all possible payoff distributions $x$, we call $\theta(x)$ as the Nucleolus. Note that the nucleolus always exists and lies in the core (if the core is non-empty), and it is unique ~\cite{schmeidler1969nucleolus}.

\subsubsection*{\textbf{Markov Core}}
In cooperative game theory, the Core ~\cite{shapley1971assignment} is the set of feasible allocations or imputations where no coalition of agents can benefit by breaking away from the grand coalition. For MDP, \cite{wang2022shaq} extended the definition of the core in cooperative game theory, called Markov core. It can assess whether the payoff distribution during the learning process in an MDP meets certain conditions, specifically that the current payoff prevents all agents from having the incentive to leave the current coalition and form new coalitions to obtain greater rewards. Formally, we modify the Markov core (MC) for our defined EC-POMDP in Subsection 3.1 as follows:
\begin{equation}\footnotesize MC =\big\{\big(\max_{a_i} x_i(s,a_i)\big)_{i \in N}\big| \max_{a_C} x(s,a_C|C) \geq \max_{a_C} v(s,a_C), \forall C \subseteq N, \forall s\in S\big\}\end{equation}
where $x_i(s,a_i)$ represent the imputation of agent $i$ by given state $s$ and action $a_i$, $x(s,a_C|C) = \sum_{i \in C}x_i(s,a_i)$ is the imputation of coalition $C$ by given state $s$ and joint action $a_C\in A_C=\times_{i\in C} A_i$, and $v(s,a_C)\in \vmathbb{R}^{+}_0$ is the assets  of coalition $C$ by given state $s$ and coalition joint action $a_C$.

\subsubsection*{\textbf{Markov Nucleolus}}
Based on the nucleolus in the cooperative game and the Markov core, we propose the Markov nucleolus to describe the credit assignment process in MARL. In EC-POMDP, $v(s, a_{CS})$ is the total assets under state $s$ and joint action $a$ while $CS$ is the coalition structure. For convience, we shorten $v(s,a_{CS})$ as $v(s,a)$. $I(v(s, a))$ represents all possible payoff distributions from the total assets to the individual payoff and each $x(s, a) = (x_1(s,a_1),\\\dots,x_n(s,a_n))\in I(v(s, a))$ is one of payoff distribution. For any coalition $C$, the unity function is defined as $u(s,a|C) =\max\limits_{a_C} v(s, a_C)$. So the excess under $x(s,a)$ is defined as follows:
\begin{align}\label{excess-mn}
e(C,x(s,a)) &= u(s, a|C) -x(s, a|C)\notag\\
&=\max_{a_C} v(s,a_C)-\sum_{i \in C} x_i(s,a_i)
\end{align}
Next, we sort the excesses of agents $N$ with $2^{n}$ coalitions under $x(s, a)$  by the non-increment as excess sequence: 
\begin{equation}\label{excess-es}\theta(x(s, a))=[e(C_1,x(s,a)),\dots, e(C_{2^n},x(s,a))]  
\end{equation}
Therefore, we can formally define the Markov Nucleolus for EC-POMDP as follows:
\begin{definition}\label{MN} For EC-POMDP, the Markov Nuclelus $x^*(s, a)\in I(v(s, a))$ is the payoff distribution that satisfied with:
\begin{equation}x^{*}(s,a) = \{\theta(x(s,a))\preceq\theta(y(s,a))\mid \forall y(s,a) \in I(v(s,a))\}
\end{equation}
which is the smallest lexicographical order.
\end{definition}

Note that the definition of the Markov nucleolus is a natural extension of the original concept from static cooperative games to MARL. We assume the Markov Nucleolus inherits the property that a nucleolus owns: there is always a unique solution, and if the Markov core is non-empty, the solution falls in the Markov core.

\subsubsection*{\textbf{Nucleolus Q-value}}We use the Markov nucleolus above to assign the global Q-value to individual Q-values. In EC-POMDP, the global Q-value under a coalition structure $CS$ is defined as: 
\begin{equation}\small
Q_{CS, global}(s, a)=R+\vmathbb{E}\left[\max_{CS'}\sum_{C \in CS'}\sum_{i \in C}Q_{i}(s',a'_i)\Bigg \vert {s' \sim P(\cdot |s,a) \atop a'_i = \max_{a_i} Q_i(s', a_i)}  \right]
\end{equation}
Next, we propose a Nucleolus Q-value based on the Markov nucleolus as follows:

\begin{corollary}\label{coro1}
To assign the global Q-value $Q_{CS,global}(s,a)$ by given state $s$ and joint action $a$ under coalition structure $CS$, we modify the payoff distribution as:
\begin{align}
&Q_{CS}(s,a) = \big \{Q_{CS,1}(s,a_1),\dots,Q_{CS,n}(s,a_n)\big \}
\end{align}
where $Q_{CS,i}(s,a_i)$ is the individual Q-value of agent $i$ and $\sum\limits_{i\in N} Q_{CS,i}(s,a_i)\\=Q_{CS, global}(s, a)$. Then, we model the Eq \ref{excess-mn} to define the excess in coalition $C$ in Q-learning. 
\begin{equation}
    e(C,Q_{CS}(s,a)) = \max_{a_C}V(s,a_C) -\sum_{i \in C }Q_{CS,i}(s,a_i)
\end{equation}
where $V(s,a_C)$ is the Q-value of coalition $C$ with given state $s$ and joint action $a_C \in A = \times_{i \in C}A_i$ and $V(s,a_C) \in \vmathbb{R}^{+}_{0}$.
We use a similar definition (Eq \ref{excess-es}) of the excess sequence in Markov nucleolus as:
\begin{equation}
    \theta(Q_{CS}(s, a))=[e(C_1,Q_{CS}(s,a)),e(C_{2^n},Q_{CS}(s,a))]
\end{equation}
Then the Nucleolus Q-value $Q^{*}_{CS}(s,a)$ can be formally defined as:
\begin{equation}\footnotesize Q^{*}_{CS}(s,a) =\{\theta(Q^*_{CS}(s,a))\preceq \theta(Q_{CS}(s,a))|\forall Q_{CS}(s,a)\in I(Q_{CS}(s,a))\}   
\end{equation}
where $I(Q_{CS.global}(s,a))$ is all possible payoff distribution coalition structure $CS$.
\end{corollary}
Further, we suppose that $Q^{*}_{CS}(s,a) = w(s,a)Q(s,a)$, where $w(s,a)$ is a vector consisted as $[w_i(s,a)]^{\top}_{i \in N}$. Based on Bellman’s optimality equation, we derive the Bellman optimality equation for the Markov nucleolus as follows:

\begin{align}\label{beleq}
Q^{*}_{CS}(s, a) =& w(s, a)\sum_{s'\in S}P(s'|s,a)[R+\gamma\max_{CS,a'} Q_{CS}(s',a')] \notag\\
= &w(s,a)\sum_{s' \in S} P(s'|s,a)[R+\gamma\max_{CS,a'}\sum_{C \in CS}\sum_{i \in C}Q^{*}_{CS,i}(s',a'_i)]
\end{align}  

Using the nucleolus-optimal Bellman equation, we derive the optimal nucleolus Q-value. To ensure the coalition's stability, we need to guarantee that the optimal nucleolus Q-value prevents agents from gaining a higher Q-value by leaving the current coalition to form a new one. And through Eq \ref{beleq}, we find that it is necessary to simultaneously optimize both the coalition structure and the action tuple $<CS, a>$ to find the optimal nucleolus Q-value. But this can result in exponential growth in the search space for the Q-values, which typically causes Q-learning to fail. To address these two issues, we derive Theorem \ref{thm-0} as follows:
%\begin{framed}
\begin{theorem}\label{thm-0}
The Nucleolus Q-value in EC-POMDP can guarantee 
\begin{enumerate}
 \item Each agent's individual Q-value in the optimal coalition structure is greater than it is in other coalition structures;
 \item The actions and the coalition structure exhibit consistency, meaning that the coalition formed under the optimal actions is also optimal.
\end{enumerate} 
\end{theorem}
%\end{framed}
The detailed proof for Theorem 1 is provided in \textbf{Appendix A.1}. Consequently, the optimal nucleolus-based Bellman equation can be reformulated as follows:
\begin{equation}\label{bellman}Q^{*}_{CS}(s,a)=w(s,a)\sum_{s' \in S}P(s'|s,a)[R+\gamma\max_{a'} \sum_{i \in N} Q^{*}_{CS',i}(s',a')]\end{equation}

\subsubsection*{\textbf{Nucleolus Q Operator}}
We construct a constrained MARL Q operator to obtain the Nucleolus Q-value and provide a theorem showing that this operator can help us achieve the optimal Nucleolus Q-value. According to the definition of the Markov Nucleolus (Subsection 3.1), we need to minimize the maximum excess value. Inspired by Reward-Constrained Policy Optimization (RCPO) ~\cite{tessler2018reward} and Discounted Reward Risk-Sensitive Actor-Critic ~\cite{prashanth2016variance}, we introduce the maximum excess as a constraint $\xi(s, a)$ in the Q-learning process. 
\begin{equation}\xi(s, a) =\max_{a_C} [V(s,a_C) - \sum_{i \in C}Q_{CS,i}^{*}(s,a_i)]
\end{equation}
Further, we construct the Lagrange multipliers as follows:
\begin{equation}\label{lag} L(\lambda,a) =\min_{\lambda \geq 0} \max_{a} [\sum_{i \in CS} Q^{*}_{CS, i}(s, a_i) + \lambda \xi(s, a)]
\end{equation}
According to RCPO, we need to implement a two-time-scale approach. This requires keeping $\lambda$ constant on the faster time scale, and optimizing policy to fix the current policy on the slower time scale and optimize $\lambda$. This process allows us to identify the saddle point of Eq \ref{lag}, which provides a feasible solution. Then, we propose an optimization operator, i.e., $\mathcal{H}: \times_{i\in N} Q^{*}_{CS,i}(s,a_i)\rightarrow  \times_{i\in N} Q^{*}_{CS, i}(s, a_i)$, with nucleolus constraints as follows:

\begin{align}
   \mathcal{H}(\times_{i \in N} Q_i^{*}(s, a_i)) 
    =&w(s,a)\sum_{s' \in S}P(s'|s,a) \notag\\
    &[R+\gamma\max_{a'} \sum_{i \in CS'} Q^{*}_{CS', i}(s, a'_i)+\lambda\xi(s, a)]
\end{align}  

Therefore, we demonstrate that using this new operator, Q learning can converge to the optimal Nucleolus Q-value by Theorem \ref{thm-2} below, where the detailed proof is provided in \textbf{Appendix A.2}.

\begin{theorem}\label{thm-2}Nucleolus-based Bellman operator can converge the optimal Nucleolus Q-value and the corresponding optimal joint deterministic policy when $\sum_{i \in N}\max_{a_i}w_i(s,a_i) \leq \frac{1}{\gamma+\lambda}$
\end{theorem}

Consequently, we believe nucleolus Q learning can help us find the maximum social welfare based on nucleolus allocation. We can conclude that no agent has an incentive to deviate from the current coalition, and this allocation provides an explanation for the credit assignment of global rewards in cooperative MARL. %Accordingly, nucleolus Q learning has the following three properties.
%\begin{itemize}
   % \item  \textit{Individual rationality}: The nucleolus ensures that each agent receives a reward no less than what they would get by acting independently. In other words, individual rationality guarantees that no one has an incentive to leave the small coalition in cooperative MARL, as they would not be better off on their own;
  %  \item \textit{Symmetry}: If agents are symmetric in the sense that they have the same role or contribution in cooperative MARL, the nucleolus will allocate the same reward to these symmetric agents. Symmetry ensures that all agents with identical conditions receive equal feedback.
  %  \item \textit{Fairness}: The nucleolus has the natural fairness merit because it prioritizes minimizing the dissatisfaction of the most discontented agents. The nucleolus achieves a globally stable and fair allocation by managing the grievances of the agent with the highest dissatisfaction and gradually reducing the dissatisfaction of all agents in cooperative MARL.
%\end{itemize}

\subsubsection*{\textbf{Implementation in Practice}}
We replace the global state $s$ in individual Q network $Q_i(s,a_i)$ with the history of local observations $\tau_i \in \tau$ for agent $i$, resulting in $Q_{i}(\tau_i,a_i)$. We handle the sequence of local observations $\tau_i$ using a recurrent neural network (RNN) ~\cite{sherstinsky2020fundamentals}. Thus, the proposed nucleolus Q-learning from Eq \ref{bellman} can be rewritten as: 
\begin{equation}
    Q^{*}_{CS}(\tau,a)=w(s,a)\sum_{s' \in S}P(s'|s,a)[R+\max_{a'} \sum_{i \in N} Q^{*}_{CS',i}(\tau_i',a'_i)]
\end{equation}
In MARL, the coalition utility function is usually unknown and must be learned through exploration. We designed our utility function network using a multi-Layer perceptron network ~\cite{rosenblatt1958perceptron} based on the critic function network in RCPO. In RCPO, the temporal difference error of the utility function network updates is as follows.
\begin{align}\label{updateV}
    \Delta_V(s,a,s') =&R+\lambda\bigg[\max_{a_C}[V(s,a_C;\phi^{-}) -\sum_{i \in C}Q^{*}_{CS,i}(s,a_i)]\bigg]\notag\\&+\eta_1\max_{a'_C}\sum_{C \in CS'}V(s',a'_C;\phi^{-})-\sum_{C \in CS} V(s,a_C;\phi)
\end{align}
where $\eta_1$ is the discounting rate of the utility function network. Similar to RCPO, when updating the utility function network, we treat the constraints $\xi(s, a)$ as part of the reward. This indicates the constraint term involving the utility network is not updated, so we use the target network $V(\cdot;\phi^{-})$ instead.

Unlike the update of the utility function network, the update of the Q network is performed using the Lagrange multiplier method. This implies that we need to optimize the Q network within the constraint conditions. Based on Eq \ref{lag}, the update of network parameters $\omega$, including both the individual Q network parameters $\varphi$ and the hypernetwork parameters $\psi$ shows as follows:
\begin{align}\label{updateQ}
    \nabla_\omega L(\lambda,\omega)=&\min_{\omega}\vmathbb{E}\bigg[R+\lambda\big[\max_{a_C}[V(s,a_C) -\sum_{i \in C}Q^{*}_{CS,i}(\tau_i,a_i;\varphi_i)]\big]\notag\\&+\gamma\sum_{i \in CS'}\max_{a'}[w(s',a';\psi^{-})Q^{*}_{CS'}(\tau',a';\varphi^{-}_i)]\notag\\&-\sum_{i \in CS} w(s,a;\psi)Q^{*}_{CS}(\tau,a;\varphi_i)\bigg]
\end{align}
where $\varphi^{-}$ represents the target Q network parameters, and $\psi^{-}$ represents the target hypernetwork parameters.
The update of multiplier $\lambda$ is shown as follows:
\begin{equation}\label{updateLA}
     \nabla_\lambda L(\lambda,\omega) = \vmathbb{E}[\lambda -\eta_2\xi(s,a)] 
\end{equation}
where $\eta_2$ is the learning rate of the update of $\lambda$.

For any coalition $C$, the input to the unity network consists of the current global state $s$ and the actions of all agents within the coalition. The network's output is the estimated value for this state-action pair. The input actions are provided by the individual Q network of each agent. To ensure a consistent input length, we use agent IDs for position encoding, placing each agent's action in the corresponding position based on its ID. For agents that are not part of the coalition $C$, we use placeholders to fill in the missing action inputs, maintaining a uniform input structure in different coalitions. In addition, we present a pseudo-code for our proposed nucleolus-based credit assignment as Algorithm 1, which provides a step-by-step flow of the learning process.
\begin{algorithm}
\caption{Nucleolus-based Credit Assignment Algorithm}\label{algo1}
\begin{algorithmic}[1]
    \State Initialize parameter $\varphi$ for agent Q network and target parameter $\varphi^-$ by copying parameter $\varphi$;
    \State Initialize parameter $\phi$ for coalition unity network and target parameter $\phi^-$ by copying parameter $\phi$  ;
    \State Initialize multiplier $\lambda \geq 0$, learning rate $\eta_1 > \eta_2 >\eta_3$ and discounting rate $\gamma$;
    \State Initialize replay buffer $\mathcal{B}$;
    \While {$Q(o_i,a_i;\omega_i)$ network not converge}
        \For{$Time \; t=1,2,\ldots to \; T$}
            \State Observe a global state $s$ and agent’s observation $o_i$;
            \State Select action $a_i$ according to each agent's Q network;
            \State Execute action $a$ and get next state $s'$, each agent's next \Statex \quad \quad \quad observation $o'_i$ and reward $R$;
            \State Store $<s,o_i \in o, s', a, R>$ to $\mathcal{B}$;
        \EndFor
        \State Sample K episodes from $\mathcal{B}$;
        \For {$Episode \; k =1,2,\ldots to \; D$}
            \For{$Time \quad t = 1,2,...,T$}
                 \State Get next time step action $a'$ by $\max_{a'} Q(s',a')$;
                 \State \textbf{Coalition unity network update}:
                 \Statex \qquad \qquad \qquad
                 $\phi_{t+1} \leftarrow \phi_{t} +\eta_1\nabla_{\phi_t}[R + \lambda\xi(s,a;\phi^{-}) $
                  \Statex \ \qquad \qquad \quad \qquad \qquad $+ \gamma V(s',a';\phi^{-})-V(s,a;\phi_{t})]$; 
                 %\tcp*[r]{Eq \ref{updateV}}
                 \Comment{ Eq \ref{updateV}}
                 \State \textbf{Nucleolus Q network update}:
                 \Statex \qquad\qquad \qquad $\varphi_{t+1} \leftarrow \varphi_t + \eta_2\nabla_{\varphi_t}[R + \lambda\xi(s,a;\varphi_t)$
                 \Statex \ \qquad \qquad \quad \qquad \qquad $ + \gamma Q(s',a';\varphi^{-})-Q(s,a;\varphi_t)]$;   
            \Comment{Eq \ref{updateQ}}
             \EndFor
            \State \textbf{Lagrange multiplier update}: 
            \Statex \qquad \qquad \quad$\lambda \leftarrow \lambda -\eta_3\xi(s,a)$; \Comment{Eq \ref{updateLA}}
            \State Update all target parameters by copying; 
        \EndFor
    \EndWhile 
 \end{algorithmic} 
\end{algorithm}

\section{Experimental Setup}
We validate our algorithm on two popular MARL benchmarks: Predator-Prey \cite{bohmer2020deep} and the StarCraft Multi-Agent Challenge (SMAC) \cite{samvelyan19smac}, against four well-known baselines, including VDN ~\cite{sunehag2017value}, QMIX ~\cite{rashid2020monotonic}, WQMIX ~\cite{rashid2020weighted}, and SHAQ ~\cite{wang2022shaq}. All baseline implementations are derived from PyMARL2 ~\cite{hu2023rethinking}.

\subsubsection*{Predator-Prey}

\begin{figure}[htbp]
    \centering
    \includegraphics[width=0.90\linewidth]{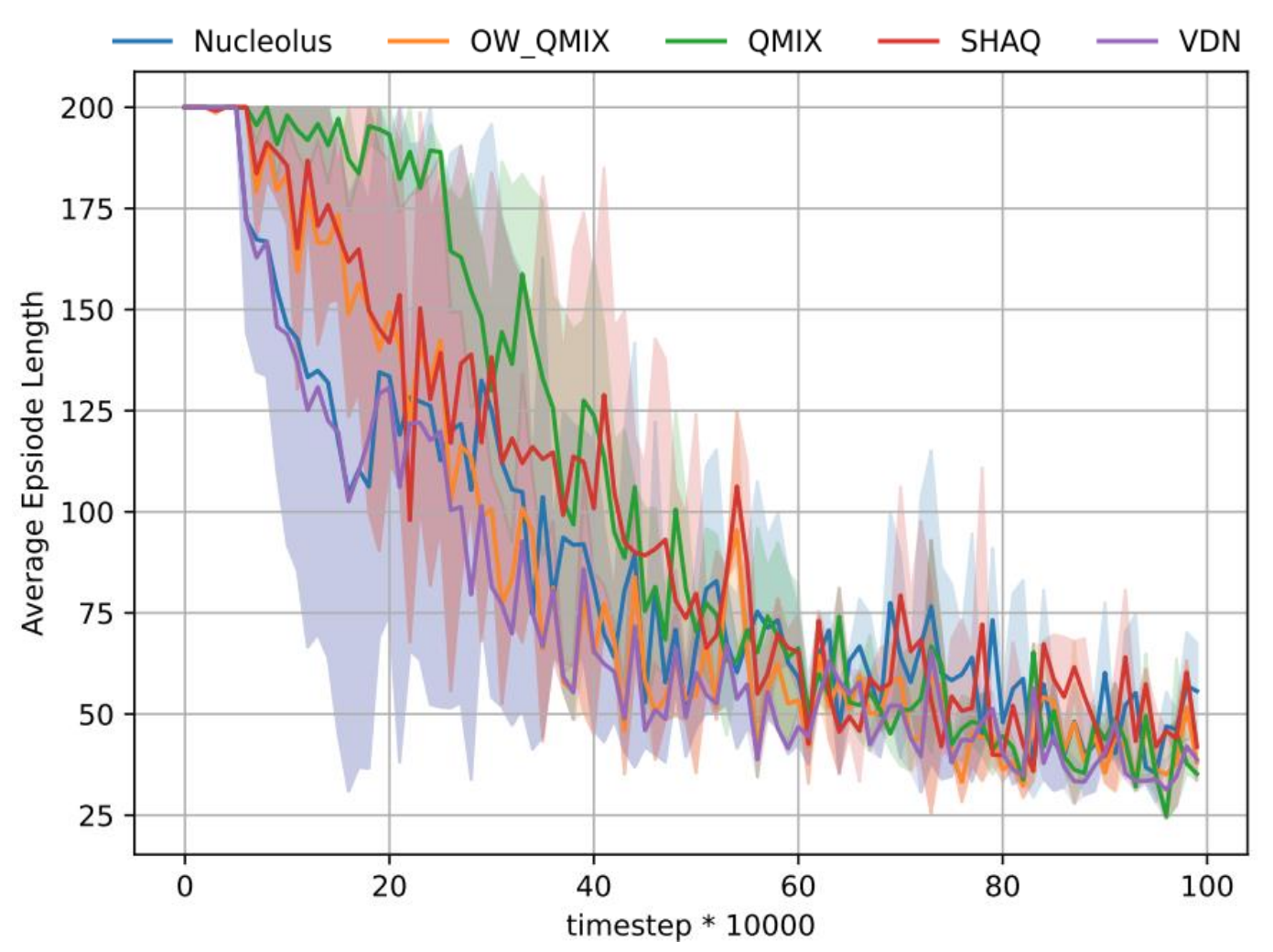}
    \caption{Learning performance in Predator-Prey: the turns to catch prey on the test episode}
    \label{pp}
\end{figure}

The Predator-Prey environment is a testbed simulating the interaction between predators and prey. Predators must cooperate to capture prey, while the prey attempts to evade capture and maximize their survival time. Each agent operates based on local observations, with predators compensating for their individual limitations through cooperation and prey using strategic movement to escape. Predators are rewarded for successful captures, which require at least two predators to capture a prey simultaneously. If the prey is captured, a global reward of 10 points is granted. The primary mission for each predator is to minimize the number of steps required to capture prey by coordinating effectively. The state space for each predator includes its position, velocity, the relative displacement between itself and the prey or other predators, and the velocity of the prey. We set 8 Predators and 4 Prey in this task, and the training duration in the Predator-Prey environment is set to 1 million steps for all algorithms.

\subsubsection*{StarCraft Multi-Agent Challenge}
SMAC is a widely used benchmark for evaluating MARL algorithms. In SMAC, each unit is controlled by an RL agent, which dynamically adapts its behavior based on local observations. Each agent is tasked with learning cooperative strategies to optimize performance in combat scenarios. Agents have a limited line of sight to perceive their local environment and can only attack enemy units within their attack range. The objective is to maximize cumulative rewards, which are calculated based on the damage inflicted on enemy units and the number of enemy units destroyed, with substantial rewards awarded for achieving overall victory. In our experiment, the training duration varies from 500k to 2.5 million steps, depending on the task difficulty. Algorithm performance is evaluated every 10k steps across 32 episodes. Detailed information on all hyperparameters of the network can be found in \textbf{Appendix B}, and descriptions of each task are present in \textbf{Appendix C}. The source code of our proposed algorithm can be reviewed in \textbf{Appendix D}, and the details of the computing platform are described in \textbf{Appendix E}.

\section{Results and Analysis}
\begin{figure*}[htbp]
    \centering
\includegraphics[width=0.9\linewidth]{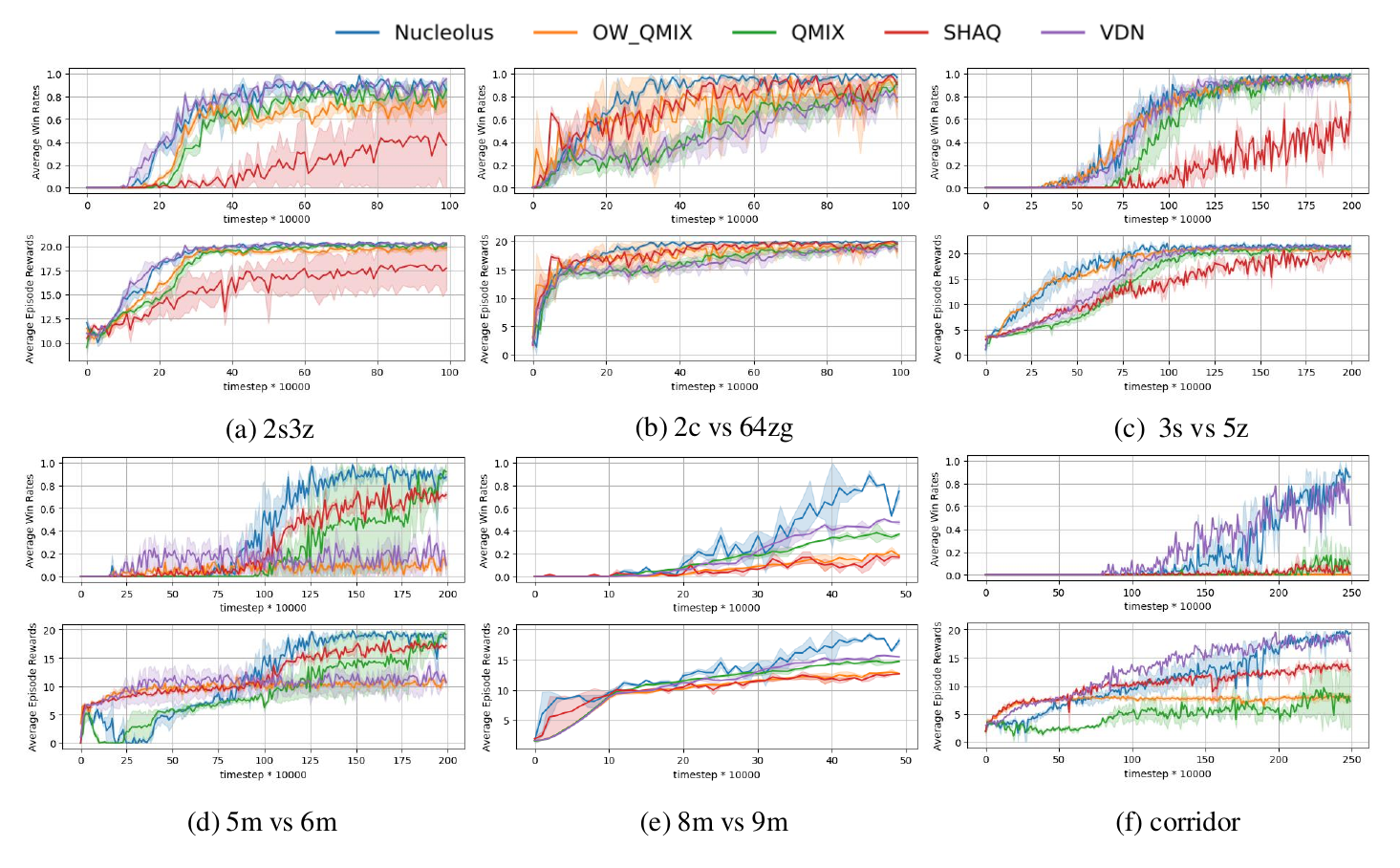}
   \caption{Learning performance in SMAC: median test win and rewards for easy task (a), hard (b-e) and super-hard (f) maps.}
\label{all performance}
\end{figure*}
In this section, we present the performance of our algorithm compared to baseline methods across two different environments: Predator-Prey and SMAC. We further analyze the coalition formation process across several tasks, highlighting how our method improves agents' cooperation in MARL.

\subsection{Learning Performance}

\subsubsection*{Predator-Prey}

\textbf{Fig.~\ref{pp}} shows the comparative performance of the five algorithms in the Predator-Prey environment, focusing on convergence speed, stable performance, and variance. VDN achieves quick convergence with low variance, yielding stable outcomes, but its overall performance is moderate. QMIX, while slower to converge, demonstrates the best final performance after 95,000 steps, capturing the prey in approximately 25 turns, though with greater fluctuations. SHAQ also exhibits slower convergence and larger performance variances.

In contrast, our proposed algorithm (nucleolus) combines rapid convergence with highly efficient capture strategies, significantly reducing the number of steps needed for successful captures. While it shows relatively higher variance, the speed and effectiveness in learning make it particularly advantageous for time-sensitive tasks. This positions our method as a strong contender, especially in scenarios demanding fast adaptation and decision-making.

\subsubsection*{SMAC}

The training curves for all algorithms in SMAC are shown in \textbf{Fig.~\ref{all performance}}, indicating varied performance across six tasks in various difficulty levels (easy, hard and super-hard).

\noindent \textbf{2s3z}: Our algorithm (nucleolus) gradually increases, reaching around 80\% after approximately 40,000 timesteps, and eventually approaching 100\%, showing stable and fast convergence. OW-QMIX behaves similarly to ours, ultimately also approaching 100\% but slightly lower than ours. QMIX performs slightly worse, with the final win rate stabilizing at around 80\%. SHAQ has a lower win rate throughout the training, stabilizing at around 45\% with high variance.VDN performs well, with a final win rate of about 85\%, though with slight fluctuations.
    
\noindent \textbf{2c vs 64zg}: Our algorithm (nucleolus) rises rapidly early on, reaching a win rate of over 90\% after about 40,000 timesteps and stabilizing near 100\%. OW-QMIX and SHAQ have a similar convergence speed to ours, reaching around an 80\% win rate at 40,000 timesteps and stabilizing at about 90\%, slightly lower than ours. QMIX and VDN converge the slowest, with a final win rate stabilizing around 80\%.
   
\noindent \textbf{3s vs 5z}: Our algorithm (nucleolus), OW-QMIX, and VDN perform well, with win rates rapidly increasing after around 300,000 timesteps and reaching approximately 80\% at 1,000,000 timesteps. The final win rate is close to 100\%. QMIX lags behind the other three algorithms, with a win rate reaching around 60\% after 1,000,000 timesteps, and eventually approaching 100\%. SHAQ converges the slowest, with the final win rate not exceeding 60\%.
\begin{figure*}[t]
    \centering
\includegraphics[width=0.9\linewidth]{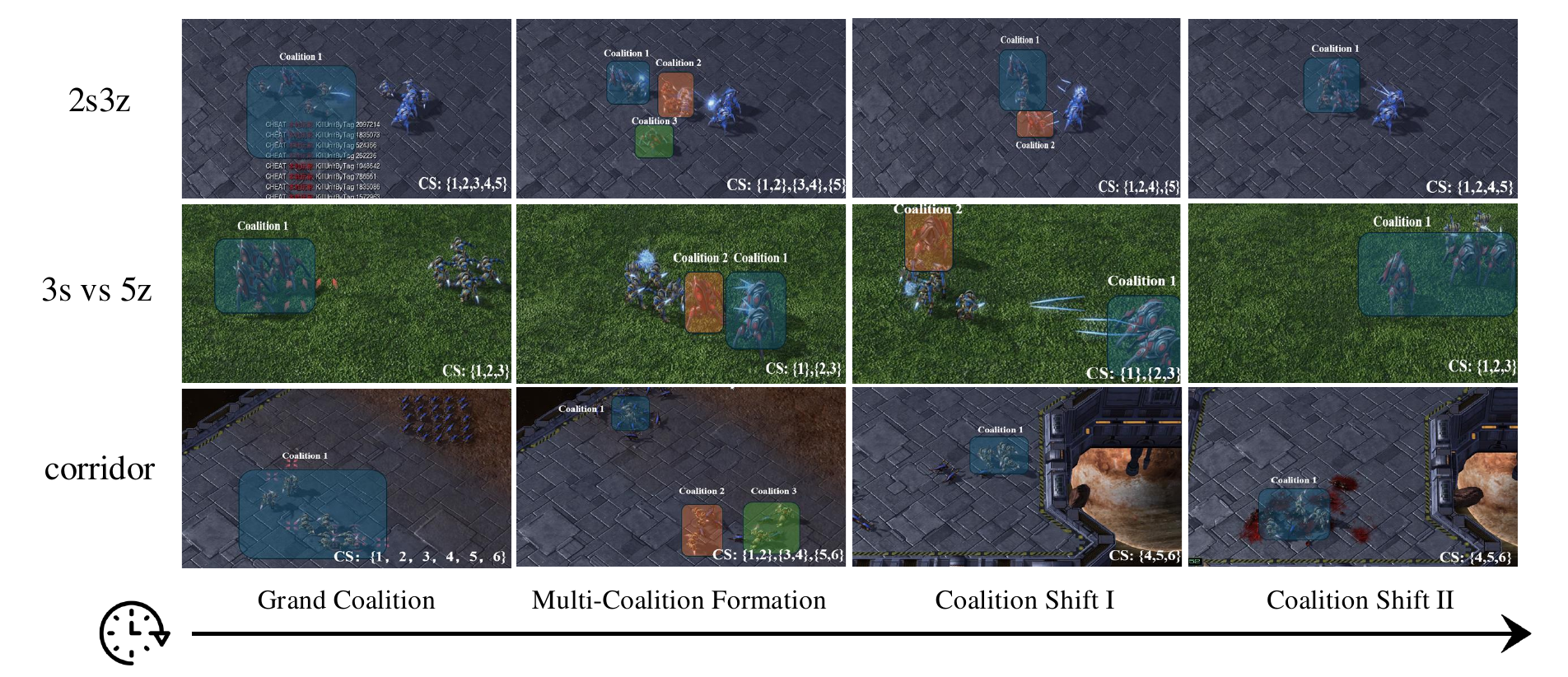}
    \caption{Visualize multiple coalition formation process in three tasks: 2s3z (easy), 3s vs 5z (hard) and corridor (super-hard)}
    \label{coalform}
\end{figure*}

\noindent \textbf{5m vs 6m}: Our algorithm (nucleolus) converges the fastest, with the win rate rapidly rising after around 100,000 timesteps and eventually approaching 100\%. OW-QMIX and VDN never exceed a win rate of 40\% throughout the training period. QMIX converges slowly and with high variance but eventually reaches a win rate close to 100\%. SHAQ converges second fastest after ours, with a final win rate stabilizing around 70\%.

\noindent \textbf{8m vs 9m}: Our algorithm (nucleolus) has the fastest convergence, with the win rate increasing rapidly early on, significantly improving after around 30,000 timesteps, and ultimately approaching 80\%. VDN converges more slowly but improves after 30,000 timesteps, with a final win rate of about 50\%. QMIX is relatively stable but increases more slowly, with the final win rate stabilizing around 40\%. SHAQ and OW-QMIX perform worse, with slow win rate improvements, eventually reaching around 20\% at 50,000 timesteps.
    
\noindent \textbf{Corridor}: Our algorithm (nucleolus) converges relatively quickly, with a final win rate exceeding 90\%, making it the best-performing algorithm. VDN has the fastest convergence, with a final win rate close to 80\%. QMIX, SHAQ, and OW-QMIX all have win rates that never exceed 20\% throughout the training period.

Based on the performance across all tasks above, our algorithm consistently achieved the highest win rates compared to the other baseline algorithms, with a superior convergence speed. Notably, in the \textbf{5m vs 6m}, \textbf{8m vs 9m}, and \textbf{corridor} tasks, our algorithm outperformed the others in terms of final win rates. We attribute these tasks are asymmetric, where the number of enemy agents significantly exceeds the number of our agents, and we found more difficult tasks require our agents to form small, dispersed alliances to succeed. In contrast, in the \textbf{2c vs 64zg} task, although the enemy agents still outnumber our agents, the smaller number of our agents makes it difficult to form effective coalitions, which limits the performance advantage compared to other algorithms. This enables our algorithm to potentially excel in handling more challenging tasks. As the number of agents increases, it can form more small coalitions, which may lead to a significantly improved performance

\subsection{Multiple Coalition Formations}

To further validate the effectiveness of our coalition formation in MARL, we visualized the coalition process on three challenging SMAC maps: 2s3z, 3svs  5z, and corridor. \textbf{Fig. \ref{coalform}} illustrates how multiple coalitions are formed and adapted to complete a composite task.

\subsubsection*{\textbf{2s3z}}

Initially, all agents belong to the \textit{Grand Coalition}. In \textit{Multi-Coalition Formation}, agents form multi-coalition, Coalition 2 (2 Zealots) draws fire on the 2 enemy Stalkers, while Coalition 1 (1 Stalker and 1 Zealot) takes the opportunity to attack the enemy Zealots, and Coalition 3 (1 Stalker) repositions, looking for a target. In \textit{Coalition Shift I}, agents shift coalition structure and form new coalitions, the remaining Stalker in Coalition 2 attacks the enemy Stalkers, while the remaining units in Coalition 1 focus fire on the enemy units other than the Stalkers. Eventually, in \textit{Coalition Shift II}, the remaining agents reform into a grand coalition to focus fire on the remaining enemy Stalkers.

\subsubsection*{\textbf{3s vs 5z}}

At the start, all agents belong to the \textit{Grand Coalition}. In \textit{Multi-Coalition Formation}, agents form multi-coalition, Coalition 2 (1 Stalker) draws all enemy fire, while Coalition 1 (2 Stalkers) takes the opportunity to attack and eliminate one enemy Zealot. In \textit{Coalition Shift I}, agents shift coalition structure and form new coalitions, the strategy is repeated, with Coalition 2 (1 Stalker) drawing enemy fire, and Coalition 1 (2 Stalkers) focusing on eliminating the remaining enemy units. As a final point, in \textit{Coalition Shift II}, the remaining agents reform into a grand coalition to focus fire on the remaining enemy Zealots.

\subsubsection*{\textbf{Corridor}}

At the beginning, all agents belong to the \textit{Grand Coalition}. In \textit{Multi-Coalition Formation}, agents form multi-coalition, Coalition 1 (2 Zealots) draws most of the enemy fire to create space for the main force to reposition, while Coalition 2 (2 Zealots) and Coalition 3 (2 Zealots) form a local numerical advantage to focus fire and eliminate smaller enemy units. In \textit{Coalition Shift I}, agents shift coalition structure and form new coalitions, the remaining agents form a grand coalition to gradually eliminate the remaining enemy units. Ultimately, in \textit{Coalition Shift II}, this strategy is repeated until all enemy units are eliminated.

\section{Conclusion}
We introduced a nucleolus-based credit assignment method to create multiple effective coalitions for cooperative MARL, leveraging cooperative game theory to enable agents to form smaller, efficient coalitions to tackle complex tasks. Unlike traditional methods that rely on a single grand coalition, our method could ensure fair and stable credit distribution across multiple coalitions. Experiments in Predator-Prey and StarCraft scenarios showed that our method consistently outperformed baseline approaches in learning speed, cumulative rewards, and win rates, demonstrating improvements in training efficiency, task performance, interpretability, and stability. Future research would focus on scaling nucleolus MARL to larger multi-agent environments, considering the computational efficiency of nucleolus calculations.

%%%%%%%%%%%%%%%%%%%%%%%%%%%%%%%%%%%%%%%%%%%%%%%%%%%%%%%%%%%%%%%%%%%%%%%%

%%% The next two lines define, first, the bibliography style to be 
%%% applied, and, second, the bibliography file to be used.

%\bibliographystyle{ACM-Reference-Format-SY} 
%\bibliographystyle{unsrt}
%\bibliography{sample}

%%%%%%%%%%%%%%%%%%%%%%%%%%%%%%%%%%%%%%%%%%%%%%%%%%%%%%%%%%%%%%%%%%%%%%%%

%%% The acknowledgments section is defined using the "acks" environment
%%% (rather than an unnumbered section). The use of this environment 
%%% ensures the proper identification of the section in the article 
%%% metadata as well as the consistent spelling of the heading.

\begin{acks}
This study is supported by Australian Research Council (ARC) DECRA Fellowship DE220100265 and ARC Discovery Project DP250103612.
\end{acks}

%%%%%%%%%%%%%%%%%%%%%%%%%%%%%%%%%%%%%%%%%%%%%%%%%%%%%%%%%%%%%%%%%%%%%%%%

%%% The next two lines define, first, the bibliography style to be 
%%% applied, and, second, the bibliography file to be used.

\bibliographystyle{ACM-Reference-Format} 
\balance
\bibliography{sample}

%%%%%%%%%%%%%%%%%%%%%%%%%%%%%%%%%%%%%%%%%%%%%%%%%%%%%%%%%%%%%%%%%%%%%%%%

\end{document}